\documentclass[preprint, preprintnumbers,nofootinbib,12pt]{revtex4-2}
\usepackage[english]{babel}
\usepackage{color}
\usepackage{amsmath}
\usepackage{physics}
\usepackage[hidelinks]{hyperref}
\usepackage{diagbox}
\usepackage{multirow}
\usepackage{amsfonts}
\usepackage{amssymb}
\usepackage{latexsym}
\usepackage{graphicx}
\usepackage{adjustbox}
\usepackage{xr}
\usepackage{siunitx}
\usepackage{threeparttable}
\usepackage{braket}
\usepackage[justification=centering]{caption}
\usepackage{float}
\usepackage{caption}
\usepackage{subcaption}
\usepackage{mathtools}
\usepackage{array}
\usepackage{seqsplit}
\usepackage{textcomp}
\usepackage{enumitem}
\usepackage{tikz}
\usepackage{tabularx}
\usepackage{setspace}
\usepackage{soul}
\usepackage{cancel}
\usepackage[scr=rsfs]{mathalpha}

\newcommand{\al}{\alpha}

\newcommand{\Si}{\Sigma}

\newcommand{\rar}{\rightarrow}

\begin{document}

\title{Potential Energy Curves of Hydrogenic Halides HX(F,Cl,Br) and 
i-DMFT Method}

\author{Horacio~Olivares-Pil\'on}
\email{horop@xanum.uam.mx}
\affiliation{Departamento de F\'isica, Universidad Aut\'onoma Metropolitana-Iztapalapa,
Apartado Postal 55-534, 09340 M\'exico, D.F., Mexico} %
\author{Alexander~V.~Turbiner}
\email{turbiner@nucleares.unam.mx}
\affiliation{Instituto de Ciencias Nucleares, Universidad Nacional
Aut\'onoma de M\'exico, Apartado Postal 70-543, 04510 M\'exico, 
D.F., Mexico}

\date{\today}

\begin{abstract}
A comparison of the {\it ab initio} calculations using the i-DMFT Method by Di Liu et al. (2025) with benchmark potential curves for three HX(F,Cl,Br) halides shows their inaccuracy in the domain around equilibrium - they do not reproduce quantitatively the results of the Born-Oppenheimer approximation - and also they predict a qualitatively wrong behavior 
in the Van der Waals region of large distances, thus, contradict the multipole expansion.
\end{abstract}

\maketitle

Recently, the ground state potential curves for diatomic molecules HF/DF/TF~\cite{AO:2022}, 
HCl/DCl/TCl~\cite{OT:2023}, HBr/DBr/TBr and HI/DI/TI~\cite{AH:2024} 
were constructed analytically with a relative accuracy $\sim 10^{-4}$ 
(or four significant digits (s.d.)). This was based on the interpolation 
of perturbation theory at small internuclear distances $R$ 
with the multipole expansion at large internuclear distances 
(Van der Waals regime) by using a two-point Pad\'e approximation
\begin{equation}
\label{Pade}
    E_d(R)\ =\ \frac{1}{R}\ \mbox{Pade}[N/N+5](R)\ \equiv \ 
    \frac{1}{R}\ \frac{P_N}{Q_{N+5}}\ ,
\end{equation}
which is ratio of two polynomials, where $N=5$ was chosen. The
parameters of the employed Pad\'e approximant $\mbox{Pade}[5/10]$ were 
fixed by requiring the exact reproduction of the first three theoretically-known 
coefficients of perturbation theory at $R \rar 0$ and the
two theoretically-known coefficients at $R \rar \infty$ expansions 
as well as three spectroscopic, experimentally-known parameters/constants. 
The eight remaining free parameters of $E_d(R)$ (1) were fixed 
by describing accurately the four low-lying vibrational energies of the 
HF, HCl, HBr and HI molecules with relative accuracy $\sim 10^{-4}$ 
(or four s.d.), respectively, via the nuclear (two-body)  
Schr\"odinger equation by adjusting the corresponding eight turning points. 
The quality of the obtained potential curves was checked by solving the 
nuclear (two-body) radial Schr\"odinger equation (without breakdown functions!)~\footnote{Hence, without modification 
of the kinetic energy, potential energy and centrifugal potential.} 
with $E_d(R)$, see (\ref{Pade}) at $N=5$, taken as a potential. 
This led to a description of {\it all} available experimental rovibrational energies 
with relative accuracy $\sim 10^{-4}$ (or four s.d.)
for {\it all} above mentioned molecules and its isotopomers collected 
in \cite{CH:2015}. 
This result was in agreement with the domain of validity of the Born-Oppenheimer approximation \cite{BO:1927}, where mass corrections $\sim m_e/M_{nuclei}$, 
QED corrections $\sim \al^2$, relativistic corrections 
$\sim v^2/c^2$ - all of the order $\lesssim 10^{-4}$ - can be neglected. 
Hence, the obtained potential curves are benchmark curves written in the form 
(1) inside of the Born-Oppenheimer approximation. It must be emphasized 
that the obtained potential curves were in agreement with the turning point 
positions, found in \cite{CH:2015}, on the level of 3-4 s.d. 
Hence, the potential energy curves (1) at $N=5$ approximate the lowest eigenvalue of the electronic 
Hamiltonian with relative accuracy $\sim 10^{-4}$.   

Solving the electronic Schr\"odinger equation in order to find the potential energy 
curves (equivalently, their eigenvalues) 
is usually the subject of the so-called {\it ab initio} calculations, 
which is a difficult problem {\it per se}, even for the two center case, 
even for a few electron case, especially in the evaluation of the rate 
of convergence of the employed method and the accuracy of obtained results. 
In fact, the only real way of evaluation (at present) is by making a comparison 
with the firmly established results obtained earlier. 

In this short Article we compare the results by Di~Liu et al. \cite{Liudi:2025}, 
see Figs. 3-4 therein~\footnote{We thank Dr.~Di~Liu for mailing us the numerical 
results on the potential curves for HF, HCl and HBr diatomic molecules, 
which were used to construct Figs 3-4.}, 
based on the recently-proposed the so-called i-DMFT method \cite{PRL:2022},
with the ground state potential curves for HF, HCl and HBr molecules 
obtained in [1-3], see Tables 1-3 below, based on the potential 
(\ref{Pade}) at $N=5$.

\begin{table}[!thb]
\caption{Electronic ground state $X^1\Si^+$ for the diatomic molecule HF: potential curve (in Hartree) 
vs. internuclear distance $R$ with the equilibrium at $R_{eq} = 1.732 54$~a.u. Second column from 
Di Liu et al.[6], see footnote 2 on p.3, third one from fit (1)~\cite{AO:2022}.}
\begin{center}
\scalebox{0.8}{%
\begin{tabular}{c| ccc}
\hline\hline
$R$ (a.u.)\ &\ Di Liu \cite{Liudi:2025}\ &\ Potential (1)~\cite{AO:2022} \\
\hline
 0.37795&  8.92077&  8.80605\\
 0.75589&  1.37778&  1.36769\\
 1.03935&  0.21252&  0.21644\\
 1.22832& -0.06327& -0.05862\\
 1.41729& -0.18064& -0.17692\\
 1.60627& -0.22104& -0.21906\\
 1.70075& -0.22557& -0.22459\\
 1.71549& -0.22564& -0.22482\\
 1.79524& -0.22374& -0.22379\\
 1.98421& -0.20844& -0.21048\\
 2.07870& -0.19744& -0.20038\\
 2.55113& -0.13478& -0.14026\\
 3.02356& -0.08239& -0.08609\\
 3.59048& -0.04338& -0.04093\\
 4.06291& -0.02694& -0.01937\\
 4.53534& -0.01875& -0.00836\\
 5.00777& -0.01444& -0.00342\\
 5.57469& -0.01151& -0.00115\\
 6.04712& -0.00994& -0.00047\\
 6.51956& -0.00880& -0.00020\\
 7.08647& -0.00779& -0.00008\\
 7.55890& -0.00713& -0.00004\\
 8.03134& -0.00659& -0.00002\\
 8.50377& -0.00611& -0.00001\\
 9.07069& -0.00560& -6.53E(-6)\\
10.01555& -0.00487& -3.39E(-6)\\
12.09425& -0.00374& -1.30E(-6)\\
14.07846& -0.00307& -6.19E(-7)\\
16.06267& -0.00257& -3.16E(-7)\\
18.89726& -0.00206& -1.33E(-7)\\
37.79452& -0.00073& -2.52E(-9)\\
56.69178& -0.00032& -2.28E(-10)\\
75.58905& -0.00012& -4.11E(-11)\\
\hline\hline
\end{tabular}}
\end{center}
\label{TcompVHF}
\end{table}

\begin{table}[!thb]
\caption{Electronic ground state $X^1\Si^+$ of the diatomic molecule HCl: potential curve (in {\it Hartree}) 
{\it vs.} internuclear distance $R$ with the equilibrium at $R_{eq} = 2.408 542$ {\it a.u.}  
Second column from Di Liu et al.\cite{Liudi:2025}, see footnote 2 on p.3, third one from fit (1)~\cite{OT:2023}.}
\begin{center}
\scalebox{0.8}{%
\begin{tabular}{c| ccc}
\hline\hline
$R$ (a.u.)\ &\ Di Liu \cite{Liudi:2025}\ &\ Potential (1)~\cite{OT:2023} \\
\hline
 0.944863&  2.228391&   2.050960       \\
 1.133836&  1.170356&   1.093860       \\
 1.511781&  0.214895&   0.201806       \\
 2.078699& -0.143275&  -0.144284       \\
 2.408645& -0.169691&  -0.169600       \\
 3.023562& -0.132582&  -0.133642       \\
 4.062911& -0.049204&  -0.051095       \\
 5.007774& -0.013979&  -0.013794       \\
 6.047124& -0.004411&  -0.002485       \\
 7.086473& -0.002508&  -0.000504       \\
 7.464418& -0.002233&  -0.000304       \\
 8.031336& -0.001956&  -0.000154       \\
 8.598254& -0.001764&  -0.000085       \\
 9.070685& -0.001639&  -0.000055       \\
 9.448631& -0.001554&  -0.000040       \\
10.015548& -0.001446&  -0.000026       \\
10.582466& -0.001353&  -0.000018       \\
11.054898& -0.001284&  -0.000013       \\
12.094247& -0.001150&  -7.39E(-6)  \\
12.566679& -0.001096&  -5.81E(-6)  \\
13.039110& -0.001048&  -4.62E(-6)  \\
13.511542& -0.001004&  -3.71E(-6)  \\
14.078460& -0.000957&  -2.89E(-6)  \\
14.550891& -0.000920&  -2.36E(-6)  \\
15.023323& -0.000886&  -1.95E(-6)  \\
15.495754& -0.000853&  -1.62E(-6)  \\
16.062672& -0.000814&  -1.30E(-6)  \\
16.535104& -0.000784&  -1.10E(-6)  \\
17.007535& -0.000754&  -9.26E(-7)  \\
17.952398& -0.000700&  -6.71E(-7)  \\
18.897261& -0.000650&  -4.94E(-7)  \\
37.794523& -0.000225&  -7.93E(-9)  \\
56.691784& -0.000097&  -7.01E(-10) \\
75.589045& -0.000036&  -1.25E(-10) \\
\hline\hline
\end{tabular}}
\end{center}
\label{TcompVHCl}
\end{table}

\begin{figure}[h!]
    \centering
    \subfloat[]{{\includegraphics[scale=1.2]{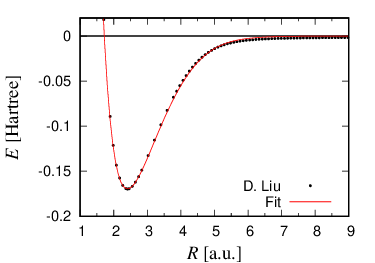} }}%
    \qquad
    \subfloat[]{{\includegraphics[scale=1.2]{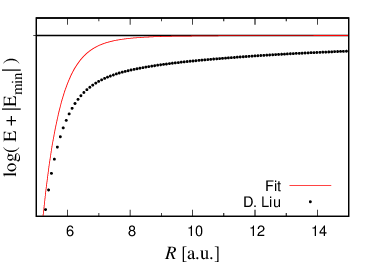} }}%
    \caption{Potential energy curve (or electronic term) for the ground state 
   $X^1\Si^+$ for the HCl molecule {\it vs.} internuclear distance $R$:
   Left panel (a) for domain $R\in [1,9]$~a.u. with data from: 
   $(i)$ Di Liu et al.\cite{Liudi:2025}, see footnote 2 on p.3 (black bullets), 
   $(ii)$ the potential curve from~\cite{OT:2023}. 
   Right panel (b) is for $R \in [5,15]$~a.u., the vertical axis given 
   by $\log{(E+|E_{min}|)}$, the horizontal line corresponds to $\log{(|E_{min}|)}$, 
   $E_{min}$ is dissociation energy.}
\end{figure}

\begin{table}[!thb]
\caption{Electronic ground state $X^1\Si^+$ of the diatomic molecule HBr: potential curve 
(in {\it Hartree})  
{\it vs.} internuclear distance $R$ with the equilibrium at $R_{eq} = 2.672 88$ {\it a.u.}
Second column from Di Liu et al.\cite{Liudi:2025}, see footnote 2 on p.3, third one from fit (1)~\cite{AH:2024}.}
\begin{center}
\scalebox{0.8}{%
\begin{tabular}{c| ccc}
\hline\hline
     $R$ ({\it a.u.})\ &\ D.Liu \cite{Liudi:2025}\ &\ Potential (1)~\cite{AH:2024}\\
\hline
 0.944863&  3.157199&   2.844940\\
 1.039349&  2.393633&   2.183360\\
 1.606267&  0.349469&   0.327671\\
 2.078699& -0.054857&  -0.059769\\
 2.456644& -0.135701&  -0.136453\\
 2.645617& -0.143933&  -0.143925\\
 3.023562& -0.132205&  -0.132035\\
 3.590480& -0.090917&  -0.091634\\
 4.062911& -0.057631&  -0.058866\\
 4.535343& -0.032510&  -0.033843\\
 5.007774& -0.016445&  -0.017690\\
 5.669178& -0.005742&  -0.006566\\
 6.047124& -0.003210&  -0.003725\\
 6.614041& -0.001516&  -0.001659\\
 7.086473& -0.000957&  -0.000891\\
 7.653391& -0.000681&  -0.000452\\
 8.031336& -0.000599&  -0.000299\\
 8.598254& -0.000537&  -0.000170\\
 9.070685& -0.000507&  -0.000110\\
 9.448631& -0.000488&  -0.000080\\
10.015548& -0.000465&  -0.000052\\
10.582466& -0.000450&  -0.000034\\
11.054898& -0.000444&  -0.000025\\
11.527329& -0.000440&  -0.000019\\
12.094247& -0.000436&  -0.000013\\
12.566679& -0.000431&  -0.000010\\
13.039110& -0.000424&  -8.06E(-6)\\
14.078460& -0.000407&  -4.85E(-6)\\
15.023323& -0.000393&  -3.18E(-6)\\
17.007535& -0.000353&  -1.44E(-6)\\
18.897261& -0.000308&  -7.47E(-7)\\
37.794523& -0.000105&  -1.09E(-8)\\
56.691784& -0.000046&  -9.49E(-10)\\
75.589045& -0.000017&  -1.68E(-10)\\
\hline\hline
\end{tabular}}
\end{center}
\label{TcompVHBr}
\end{table}

\begin{figure}[h!]
    \centering
    \subfloat[]{{\includegraphics[scale=1.2]{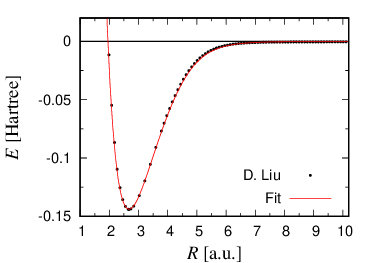} }}%
    \qquad
    \subfloat[]{{\includegraphics[scale=1.2]{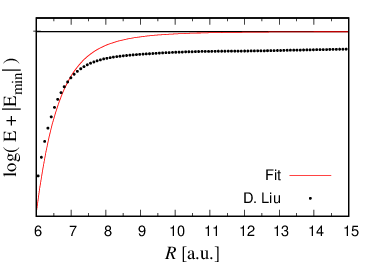} }}%
    \caption{Potential energy curve (or electronic term) for the ground state 
   $X^1\Si^+$ for the HBr molecule {\it vs.} internuclear distance $R$:
   Left panel (a) for domain $R\in [1,10]$~a.u. with 
   data from: 
   $(i)$ Di Liu et al.\cite{Liudi:2025}, see footnote 2 on p.3 (black bullets), 
   $(ii)$ the potential curve from~\cite{OT:2023}. Right panel (b) is for 
   $R\in [6,15]$~a.u., the vertical axis given 
   by $\log{(E+|E_{min}|)}$, the horizontal line corresponds to $\log{(|E_{min}|)}$, 
   $E_{min}$ is dissociation energy.}
\end{figure}

One can see that systematically the i-DMFT method does {\it not} reproduce 
four s.d. in the benchmark potential energy curves (1), thus, it deviates 
from the turning points obtained in \cite{CH:2015}, even in a domain 
in the vicinity of the minimum close to equilibrium , where there is an agreement 
in 3 s.d. The absolute deviation becomes truly significant beyond equilibrium, 
at small and especially at large internuclear distances. In particular, 
at $R > 5${\it a.u.} for the HF molecule, $R > 7${\it a.u.} for the HCl molecule 
and at $R > 9${\it a.u.} for the HBr molecule the deviation is growing fast with 
the increase of $R$ and can reach several orders of magnitude for a large $R$, 
see Tables 1-3; in particular, at $R \sim 50 - 75$~{\it a.u.} it reaches five orders 
of magnitude!
Hence, this indicates that the behavior of the potential energy curves 
obtained in \cite{Liudi:2025} in the Van der Waals regime does {\it not} agree 
with the multipole expansion. 
This implies that by taking the turning points found using the i-DMFT method 
as basic ones and placing them on some potential curve will lead to inaccurate 
rovibrational spectra, even for low-lying states. It is confirmed by making a comparison 
of the vibrational energies $E_{\nu 0}$ obtained in \cite{Liudi:2025} 
(see {\it Supporting information} therein) with the results from [1-4] 
\footnote{which are in agreement with experimental vibrational spectra in, 
at least, four s.d.}: 
one can see that for small $\nu$ three s.d. are reproduced only (as expected), 
then for larger $\nu < 17$  - one-two s.d. while for $\nu = 17 - 20$ not even single s.d. 
is reproduced! \footnote{Note that in similar manner the low accuracy results for 
vibrational energies for $F_2, Cl_2$, and $Br_2$ molecules, obtained in the i-DMFT method, 
are reported in \cite{Liudi:2024}.}
This indicates shortcomings of the i-DMFT method. The question, of
how intrinsic to the method used these shortcomings are, needs to be investigated and 
clarified before drawing conclusions about the heuristic value of the i-DMFT method if any. 

\section*{Acknowledgments}

\noindent
This work is partially supported by DGAPA grant IN113025 (Mexico).
We are grateful to J.~Karwowski and A.~Orr-Ewing, for suggestions 
which led to the improvement and the further clarity of the presentation.

\end{document}